\documentclass[12pt,a4paper,final]{iopart}

\usepackage{iopams}  
\usepackage[breaklinks=true,colorlinks=true,linkcolor=blue,urlcolor=blue,citecolor=blue]{hyperref}
\usepackage{graphicx}
\usepackage{csquotes}
\usepackage{float}
\usepackage{stackrel}
\usepackage{amssymb}
\usepackage{amsthm}
\usepackage{subfig} 
\usepackage{array}
\usepackage{algorithm}
\usepackage{algpseudocode}

\newcommand{\mt}[1]{\mathrm{#1}} 
\newcommand{\Reals}[1]{\mathbb{R}^{#1}}        
\newcommand{\neurc}{\mathrm{\boldsymbol{J}}^{p}}        
\newcommand{\Avec}{\boldsymbol{A}}        
\newcommand{\Hilbert}{\mathbb{H}}
\newcommand{\rvec}{\boldsymbol{r}}        
\newcommand{\nvec}{\boldsymbol{n}}        
\newcommand{\rvecunit}{\hat{\boldsymbol{r}}}        

\newcommand{\tvec}{\boldsymbol\tau}
\newcommand{\qvec}{\boldsymbol{Q}}
\newcommand{\tvecunit}{\hat{\boldsymbol\tau}}
\newcommand{\lambdavec}{\boldsymbol{\lambda}}
\newcommand{\bvec}{\boldsymbol{\mathrm{b}}}
\newcommand{\zv}{\boldsymbol{0}}

\newcommand{\alphaopt}{\alpha_{\textrm{opt}}}
\newcommand{\rl}{\underline{\mathbf{r}}}

\newcommand{\Exp}{\mathbb{E}}
\newcommand{\lambdaest}{\hat{\boldsymbol{\lambda}}}
\newcommand{\wgn}{\mathbf{w}}

\newcommand{\Jvec}{\mathrm{\bold{J}}^{p}}        

\newcommand{\Psiv}{\boldsymbol{\Psi}}        
\newcommand{\PsiL}{\boldsymbol{\Psi}_L}        
\newcommand{\hPsiL}{\boldsymbol{\hat{\Psi}}_L}        
\newcommand{\hPsi}{\boldsymbol{\hat{\Psi}}}        
\newcommand{\uvec}{\boldsymbol{u}}

\begin{document}

\title{A hybrid analytical-numerical algorithm for determining the neuronal current via EEG}

\author{Parham Hashemzadeh $^1$, A.~ S. Fokas $^{1,2}$,  C.~B.~ Sch\"{o}nlieb $^1$ }
\address{$^1$ Department of Applied Mathematics and Theoretical Physics, University of Cambridge, Cambridge CB3 0WA, UK. }
\address{$^2$ Department of Electrical Engineering, University of Southern California, Los Angeles, California 90089, USA.}

\ead{ph442@cam.ac.uk.}
\begin{abstract}
In this study, the neuronal current in the brain is represented using Helmholtz decomposition. It was shown in earlier work that data obtained via electroencephalography (EEG) are  affected only by the irrotational component of the current. The irrotational component is denoted by $\Psi$ and has support in the cerebrum. This inverse problem is severely ill-posed and requires that additional constraints are imposed. Here, we impose the requirement of the minimization of the $L_2$ norm of the current (energy). The function $\Psi$ is expanded in terms of inverse multiquadric radial basis functions (RBF) on a uniform Cartesian grid inside the cerebrum. The minimal energy constraint in conjunction with the RBF parametrization of $\Psi$ results in a Tikhonov regularized solution of $\Psi$. The RBF shape parameter
(regularization parameter), is computed by solving a 1-D non-linear maximization problem. Reconstructions are presented using synthetic data with a signal to noise ratio (SNR) of $20$ dB. The root mean square error (RMSE) between the exact and the reconstructed $\Psi$ is RMSE=$0.1122$. The proposed reconstruction algorithm is computationally efficient and can be vectorized in MATLAB.
\end{abstract}
%
\par
The medical significance of Electro-encephalography, EEG, is well established, see for example \cite{GCC,papanicolaou2006,Niedermeyer2011}. The estimation of the neuronal current from the measured electric potential (units Volts) on the surface of the head provided via the EEG data, can be formulated as a mathematical inverse problem \cite{kaipio2004,fokas1996,mendez2000,BML,sloreta2002,hauk2004,dassios2017,dassios2009,fokas2009,hashemzadeh2018}. The problem of computing the electric potential for a given head model and a given configuration of dipole sources is referred to as the forward problem \cite{mosher1999, kybic2005, openmeeg2010,vorkwek2012}. Solving the forward problem is a pre-requisite for the solution of the inverse EEG problem. If the head model is approximated as a volume conductor consisting of nested compartments with constant conductivities, then the forward problem can be formulated as a set of boundary integral equations \cite{mosher1999, kybic2005, openmeeg2010,openmeeg2011,vorkwek2012,Geselowitz1967, hamalainen1989}. OpenMEEG \cite{openmeeg2010,openmeeg2011} is an accurate boundary element solver (BEM) that solves these boundary integral equations. More precisely, it solves the forward problem for an arbitrary-shaped piecewise homogeneous conductor and a set of dipole sources. In this setting, the standardized low resolution brain electromagnetic tomography (sLORETA) technique \cite{sloreta2002} is known to provide accurate solutions. Other approaches which also use discrete formulations can be found in \cite{darbas2019,miinalainen2019,pursiainen2016}. In   
contrast to the above important approaches, here the neuronal current is modelled as a continuous vectorial function. This is a more accurate representation of the underlying physics. Our study concetrates on the inverse problem of reconstructing $\Psi$ from EEG measurements. For solving this inverse problem, we use a result from \cite{fokas2009} that expresses the electric potential on the scalp (measured via EEG) in terms of an integral over the cortical volume involving the product of the Laplacian of $\Psi$ and a certain auxiliary function denoted by $v_s$; the latter depends on the geometry of the various compartments and their conductivities but not on the current:
\begin{equation}
\label{2ndexpress0}
u_s(\rvec)=-\frac{1}{4\pi} \int_{\Omega_c} \nabla_{\tvec}^{2} \Psi(\tvec) v_s(\rvec,\tvec)dV(\tvec), \qquad \rvec \in S_s ,
\end{equation}
where $u_s(\rvec)$ denotes the electric potential (units Volts) measured on the scalp at position vector $\rvec$ via EEG, $\Psi(\tvec)$ is the unknown scalar function to be reconstructed (irrotational component of the current), $\Omega_c$ denotes the volume of the cerebrum and $S_s$ is the surface of the scalp.

\par
The outline of the paper is as follows: the head model considered in this study is discussed in section \ref{headM}. The measurement equation, expressing the electric potential  (units Volts) on the scalp in terms of $\Psi(\tvec)$, and the auxiliary function $v_s(\rvec,\tvec)$ is reviewed in section \ref{forwardproblem}. The minimum norm framework and the radial basis function (RBF) parametrization of $\Psi(\tvec)$ is derived in section \ref{inverseproblem}. Numerical results and reconstructions for the case of a realistic head model are presented in section \ref{numerics}. Finally, conclusions are discussed in section \ref{conclusion}.

\section{Head Model}
\label{headM}
\begin{figure}[H]
 \includegraphics[scale=0.5]{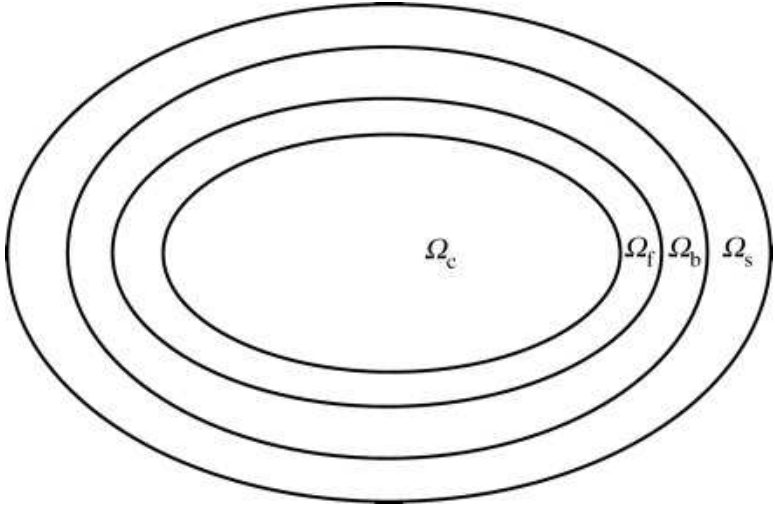}
  \caption{The different compartments of the head model. $\Omega_c$ denotes the cerebrum that is surrounded by three shells $\Omega_f$, $\Omega_b$, $\Omega_s$, denoting the cerebrospinal fluid, the skull, and the scalp. Their conductivities are respectively denoted by $\sigma_c$, $\sigma_f$, $\sigma_b$ and $\sigma_s$. The spaces $\Omega_c$, $\Omega_f$, $\Omega_b$ and $\Omega_s$ are bounded by the surfaces $S_c$, $(S_c,S_f)$, $(S_f,S_b)$ and $(S_b,S_s)$. }
  \label{headmodel}
\end{figure}

The different compartments of the head model are shown in figure \ref{headmodel}. The bounded domain $\Omega_c$ represent the cerebrum, which has conductivity $\sigma_c$. A shell $\Omega_f$ with conductivity $\sigma_f$, representing the cerebrospinal fluid, surrounds the domain $\Omega_c$. The cerebrospinal fluid that is surrounded by the skull is characterized by the domain $\Omega_b$ with conductivity $\sigma_b$. Finally, the skull is surrounded by the scalp that is modelled as a shell $\Omega_s$ with conductivity $\sigma_s$. Notations for the surfaces forming the boundaries of above domains are introduced in figure \ref{headmodel}. The domain exterior to the head is denoted by $\Omega_e$, and it is assumed that $\Omega_e$ is not conductive. The permeability of all domains are equal to the permeability $\mu$ of empty space. 

Table \ref{tabl4layer} presents the conductivity values of the head model as documented in  \cite{gabriel1996,manola2005,hoekema2003}.
\begin{table}[H]
\begin{center}
\begin{tabular}{ |c||c|c| }
 \hline
 Domain $\Omega$& Conductivity $\sigma$ (S/m) \\
 \hline
 Cerebrum $\Omega_c$& 0.33 \\
 \hline
 CSF $\Omega_f$ & 1.0 \\
 \hline
 Skull $\Omega_b$&  0.0042 \\
 \hline
 Scalp $\Omega_s$&  0.33 \\
 \hline
 \end{tabular}
 \end{center}
 \caption{The conductivity values for the different compartments of the head model.}
 \label{tabl4layer} 
 \end{table}

The physiology of the head model is accurately characterized by the four layer compartments shown in figure \ref{headmodel}, and the results derived in \cite{fokas2009} are valid in the four layer head model. It can be observed from table \ref{tabl4layer} that $\Omega_f$ (CSF), has a higher conductivity than the remaining compartments, but it also has a very small thickness. The detailed analysis of \cite{clerk2009} shows that the brain-CSF interface has a negligible effect in the analysis. For this reason, for our numerical examples we ignore $\Omega_f$ (CSF) and restrict our analysis to the three layer head model involving the compartments $\Omega_c, \Omega_b$ and $\Omega_s$.

\section{The Measurement Equation}
\label{forwardproblem}
Let $\neurc(\tvec)$, $\tvec \in \Omega_c$, denote the primary current (neuronal current) which is assumed to be supported within the cerebrum $\Omega_c$. Under the assumption that $\Jvec(\tvec)$ has sufficient smoothness (continuous derivatives), we can employ the Helmholtz decomposition to express $\Jvec$ in terms of its irrotational and solenoidal components: 
\begin{equation}
\Jvec(\tvec)=\nabla_{\tvec} \Psi(\tvec)+ \nabla_{\tvec} \times \Avec(\tvec), \qquad \tvec \in \Omega_c,
\label{helmddecomp}
\end{equation}
where $\Avec(\tvec)$ characterises  the solenoidal component and satisfies the constraint 
$\nabla \cdot \Avec(\tvec)=0$. $\Psi(\tvec)$ characterises the irrotational component. Under  the assumption that $\Jvec(\tvec)$ vanishes on the cortical surface $S_c$, it is shown in \cite{fokas2009} that the electric potential measured on the scalp, $S_s$, is given by 
\begin{equation}
\label{2ndexpress2}
u_s(\rvec)=-\frac{1}{4\pi} \int_{\Omega_c} \nabla_{\tvec}^{2} \Psi(\tvec) v_s(\rvec,\tvec)dV(\tvec), \qquad \rvec \in S_s ,
\end{equation}
where the auxiliary function $v_s(\rvec,\tvec)$ depends on geometry and conductivities but is independent of the current \cite{fokas2009}. It was shown further in \cite{hashemzadeh2018} that equation (\ref{2ndexpress2}) can be expressed in the form of a surface integral :
\begin{equation}
\label{2ndexpress4}
u_s(\rvec)=\frac{1}{4\pi} \int_{S_c} \nvec \cdot \biggl[ \Psi(\tvec) \nabla_{\tvec} v_s(\rvec,\tvec) 
-v_s(\rvec,\tvec) \nabla_{\tvec} \Psi(\tvec)\biggr]dS(\tvec), \quad \rvec \in S_s. 
\end{equation}
Thus,  $u_s(\rvec)$ is only affected  by the values of $\Psi$ and of $\nabla_{\tvec} \Psi$ on the surface of $S_c$.  It is therefore \textit{impossible} to determine the $3$-dimensional nature of $\Psi$ from the knowledge of $u_s(\rvec)$. Assuming that $v_s(\rvec,\tvec)$  can be computed, equation (\ref{2ndexpress2}) provides a relation between $u_s(\rvec)$ and $\nabla_{\tvec}^{2}\Psi(\tvec)$. Similarly, assuming that both $v_s(\rvec,\tvec)$ and $\nvec \cdot \nabla_{\tvec}v_s(\rvec,\tvec)$ can be computed, equation (\ref{2ndexpress4}) provides a relation between $u_s(\rvec)$, $\Psi(\tvec)$ and $\nabla_{\tvec} \Psi(\tvec)$.

\subsection{Auxiliary Functions $v_j(\rvec,\tvec)$}
\label{compvs}
It was shown in \cite{fokas2009} that for a given geometry, the functions $v_j(\rvec,\tvec)$, $j=c,f,b,s$ are defined via the following boundary value problem:

\begin{eqnarray}
\label{bveqn1}
&\nabla_{\tvec}^2 v_c(\rvec,\tvec)=0,\quad  \rvec \in \Omega_c, \tvec \in \Omega_c, \\
&\frac{\partial}{\partial n}\biggl[ \frac{1}{|\rvec-\tvec|}+v_c(\rvec,\tvec) \biggr]=\sigma_f \frac{\partial v_f(\rvec,\tvec)}{\partial n}, \quad\rvec \in S_c; \nonumber
\end{eqnarray}

\begin{eqnarray}
\label{bveqn2}
&\nabla_{\tvec}^2 v_f(\rvec,\tvec)=0, \quad \rvec \in \Omega_f, \tvec \in \Omega_c,\\
&v_f(\rvec,\tvec)=\frac{1}{\sigma_c} \biggl[ \frac{1}{|\rvec-\tvec|}+v_c(\rvec,\tvec)\biggr], \qquad \rvec \in S_c, \nonumber\\ 
&\sigma_f \frac{\partial v_f(\rvec,\tvec)}{\partial n} =\sigma_b \frac{\partial v_b(\rvec,\tvec)}{\partial n},\qquad \rvec \in S_f, \tvec \in \Omega_c;\nonumber
\end{eqnarray}

\begin{eqnarray}
\label{bveqn3}
&\nabla_{\tvec}^2 v_b(\rvec,\tvec)=0, \quad \rvec \in \Omega_b, \tvec \in \Omega_c,\\
&v_b(\rvec,\tvec)=v_f(\rvec,\tvec), \quad \rvec \in S_f, \nonumber \\  
&\sigma_b \frac{\partial v_b(\rvec,\tvec)}{\partial n} =\sigma_s \frac{\partial v_s(\rvec,\tvec)}{\partial n},\qquad \rvec \in S_b, \tvec \in \Omega_c;\nonumber 
\end{eqnarray}

\begin{eqnarray}
\label{bveqn4}
&\nabla_{\tvec}^2 v_s(\rvec,\tvec)=0, \quad \rvec \in \Omega_s, \tvec \in \Omega_c,\\
&v_s(\rvec,\tvec)=v_b(\rvec,\tvec), \quad \rvec \in S_b,\nonumber\\ 
&\frac{\partial v_s(\rvec,\tvec)}{\partial n} =0, \quad \rvec \in S_s.\nonumber
\end{eqnarray}

Equations (\ref{bveqn1}) - (\ref{bveqn4}) are \textit{independent} of the current $\Jvec(\tvec)$ and depend only on the geometry and on the conductivities $\sigma_c$, $\sigma_b$, $\sigma_f$ and $\sigma_s$.\par
It is shown in \cite{fokas2009} that the functions $v_j(\rvec,\tvec)$ can be related to the functions  $u_j(\rvec,\tvec)$ with unit Volts, $\rvec \in \Omega_j, ~~\tvec \in \Omega_c, ~~j \in \{c,f,b,s \}$. These functions are defined in terms of a single dipole with moment $\qvec(\tvec)$ with unit Coulomb-meter, located at the position vector $\tvec$ via the following equations:

\begin{eqnarray}
\label{bv1}
&\sigma_c \nabla^2_{\tvec} u_c(\rvec,\tvec)=\nabla_{\tvec} \cdot \boldsymbol{Q} \delta (\rvec-\tvec), \quad \rvec \in \Omega_c,\\
&\sigma_c \frac{\partial u_c(\rvec,\tvec)}{\partial n}=\sigma_f \frac{\partial u_f(\rvec,\tvec)}{\partial n}, \quad \rvec \in S_c;\nonumber
\end{eqnarray}

\begin{eqnarray}
\label{bv2}
&\nabla_{\tvec}^2 u_f(\rvec,\tvec)=0, \quad \rvec \in \Omega_f, \\
&u_f(\rvec,\tvec)=u_c(\rvec,\tvec), \quad \rvec \in S_c,\nonumber\\
&\sigma_f \frac{\partial u_f(\rvec,\tvec)}{\partial n}=\sigma_b \frac{\partial u_b(\rvec,\tvec)}{\partial n}, \quad \rvec \in S_f;\nonumber
\end{eqnarray}
\begin{eqnarray}
\label{bv3}
&\nabla_{\tvec}^2u_b(\rvec,\tvec)=0,  \quad \rvec \in \Omega_b,\\
&u_b(\rvec,\tvec)=u_f(\rvec,\tvec), \quad \rvec \in S_f,\nonumber\\
&\sigma_b \frac{\partial u_b(\rvec,\tvec)}{\partial n}=\sigma_s \frac{\partial u_s(\rvec,\tvec)}{\partial n}, \quad \rvec \in S_b;\nonumber
\end{eqnarray}
\begin{eqnarray}
\label{bv4}
&\nabla_{\tvec}^2u_s(\rvec,\tvec)=0,  \quad \rvec \in \Omega_s,\\
&u_s(\rvec,\tvec)=u_b(\rvec,\tvec), \quad \rvec \in S_b,\nonumber\\
&\frac{\partial u_s(\rvec,\tvec)}{\partial n}=0, \quad \rvec \in S_s.\nonumber
\end{eqnarray}

The functions $u_j$ and $v_j$ are related by the equation
\begin{eqnarray}
\label{connection}
&u_j(\rvec,\tvec)=\frac{1}{4\pi}\qvec(\tvec)\cdot \nabla_{\tvec} v_j(\rvec,\tvec), \\
& j\in \{f,b,s\}, \rvec \in \Omega_j, \tvec \in \Omega_c\nonumber.
\end{eqnarray}
It was shown in \cite{dassios2017,dassios2009,fokas2009} that for the the particular case of the spherical head model, $v_s(\rvec,\tvec)$ is given by 
\begin{equation}
v_s(\rvec,\tvec)=\sum_{n=1}^{\infty} H_n \frac{2n+1}{4\pi}\frac{\tau^n}{r^{n+1}} P_n(\tvecunit \cdot \rvecunit),
\label{vsansatz}
\end{equation}
where the  coefficients $H_n$ depend only on the conductivities and the geometry and do not depend on the neuronal current. Equation (\ref{vsansatz}) shows that the auxiliary function $v_s(\rvec,\tvec)$ depends on the three variables $(\tau,r,\rvecunit \cdot \tvecunit)$. It is worth noting that in the particular case of the spherical head model, $v_s(\rvec,\bold{0})=0$. \par

In the case of a realistic head model, OpenMEEG \cite{openmeeg2010} accurately solves the boundary value problem described by equations (\ref{bv1})-(\ref{bv4}) but not equations (\ref{bveqn1})-(\ref{bveqn4}). However, we can employ \textit{the fundamental theorem of line integrals} to approximate $v_s(\rvec,\tvec)$ from solutions of $u_s(\rvec,\tvec)=\frac{1}{4\pi}\qvec \cdot \nabla_{\tvec}v_s(\rvec,\tvec)$ computed by OpenMEEG. \par
For completeness, we recall the \textit{fundamental theorem of line integrals}. Suppose that $C$ is a smooth curve given by $\rl(t)$, $a \leq t \leq b$ and let $f(\rl(t))$ be a function whose gradient $\nabla f(\rl(t))$ is continuous on $C$. Then, 
\begin{equation}
f(\rl(b))-f(\rl(a))=\int_{C} \nabla f(\rl(t)) \cdot d\rl(t),
\label{fundlineinteg}
\end{equation}
where $\rl(a)$ and $\rl(b)$ represent the initial and final points on $C$, respectively (the above theorem holds in any number of dimensions). \par
If we let $f=v_s(\rvec,\tvec)$ and $\rl(t)=\tvecunit t$, $0\leq t\leq \tau$ in equation (\ref{fundlineinteg}), then we find
\begin{equation}
v_s(\rvec,\tvec)-v_s(\rvec,\bold{0})=4\pi\int_{0}^{\tau} u_s(\rvec,t \tvecunit,\tvecunit) dt,
\label{vslineinteg}
\end{equation}
where $v_s(\rvec,\bold{0})$ is a constant and $4\pi u_s(\rvec,\tvec,\tvecunit)=\tvecunit \cdot \nabla_{\tvec}v_s(\rvec,\tvec)$. It is straightforward to compute numerically  the one-dimensional integral of equation (\ref{vslineinteg}). In the case of the spherical head model, $v_s(\rvec,\bold{0})=0$ (see equation (\ref{vsansatz})). If the centroid of the brain mesh of a realistic head model is denoted by $\bold{c}$, then $v_s(\rvec,\bold{c})$ is an additive constant and has no effect on the inversion formulas of equations (\ref{2ndexpress2}) and (\ref{2ndexpress4}). If we estimate  $\bold{c}$ from the triangulated surface mesh of the brain by fitting a sphere to its nodes, then a reasonable assumption is $v_s(\rvec,\bold{c})\approx 0$. The line integral of equation (\ref{vslineinteg}) has to be computed numerically.  \par
The accuracy of reconstructing $\Psi(\tvec)$ using the volume integral of equation (\ref{2ndexpress2}) is dependent on $v_s(\rvec,\tvec)$. One approach is to employ a large number of sources $\qvec(\tvec)$, to generate the required data set $u_s(\rvec,\tvec,\tvecunit)$ and to approximate the line integral of equation (\ref{vslineinteg}), using a simplified integration method. However, this approach \textit{might be} computationally expensive and requires certain level of memory management. Furthermore, it is also well known that sources placed too close to the surface of cerebrum $S_c$, can result in large numerical errors in $u_s(\rvec,\tvec,\tvecunit)$, see \cite{openmeeg2010,openmeeg2011}. This in turn, will introduce errors in the upper limits of the line integral of equation (\ref{vslineinteg}). A computationally tractable alternative, that  
can mitigate these errors involves  \textit{a carefully} constructed regression model of $u_s(\rvec,\tvec,\tvecunit)$ using data generated via OpenMEEG. The details of this construction are discussed in section \ref{surrogatemodel}.

\subsection{Regression Model of $u_s(\rvec,\tvec,\tvecunit)$ For A Realistic Head Model}
\label{surrogatemodel}
Regression models constructed using data generated by PDE solvers are referred to as \textit{surrogate models} \cite{forrester2008,simpson2008,forrester2009,grihon2013, pruett2016,raul2018}. They are essentially machine learning models, that \textit{approximate} a mapping between a set of inputs and outputs of a simulation. Suppose we are given a finite sample pairs of data, $(\bold{x}_n,\bold{y}_n)_{n=1}^{N}$ (the training data), where $\bold{x} \in \Reals{d_{in}}$ and $\bold{y} \in \Reals{d_{out}}$. These pairs of data represent the inputs and outputs of a computationally expensive simulation model $\bold{y}=f(\bold{x})$. Our goal is to construct a function $\hat{f}: \Reals{d_{in}} \rightarrow \Reals{d_{out}}$, which is as close as possible to the true function $f$. In this study, function $f=u_s(\rvec,t \tvecunit,\tvecunit)$ on the right hand side of equation (\ref{vslineinteg}). A vital step in constructing an accurate surrogate model is the correct identification of the inputs to the model. In this connection we use our insight from the analysis of the spherical head model: equation (\ref{vsansatz}) shows that $v_s(\rvec,\tvec)$ is a function of $r$, which is the radial distance from the center of the coordinate system to the sensor position of $\tau$, which is the radial distance from the center of the coordinate system to the source position vector, and of $(\rvecunit \cdot \tvecunit)$, which is the cosine of the angle between the source and observation unit vectors.  Thus, the necessary inputs to the surrogate model are $(r,\tau, \rvecunit\cdot \tvecunit)$. An accurate \textit{learning} of the function $u_s(\rvec,t \tvecunit,\tvecunit)$  depends on a dense sampling of the inputs $(r,\tau, \rvecunit\cdot \tvecunit) \in [r_{min},r_{max}]\times [\tau_{min},\tau_{max}] \times [-1,1]$. The position vectors of the observation points and source points from the centroid $\bold{c}$ of the brain are denoted by  $\rvec$ and $\tvec$ respectively. The position vectors of the nodes of the brain with respect to $(0,0,0)$ are denoted by $\{\bold{x}_j \in \Reals{3}: 1\leq j\leq V_b\}$; similarly the position vectors of the nodes of the scalp are denoted by $\{\bold{y}_j \in \Reals{3}:1\leq j\leq V_s\}$. Recall that OpenMEEG solves the boundary value problem described by equations (\ref{bv1})-(\ref{bv4}). For any source-observation  pair $(\tvec,\rvec)$, OpenMEEG employs a current dipole source with moment $\qvec(\tvec)$ Coulomb-meter. This solution, which has units Volts, will be denoted by $u_s(\rvec,\tvec,\qvec(\tvec))$. The steps needed for generating the data points required for constructing the surrogate model are given below:

\begin{algorithm}
 \renewcommand{\thealgorithm}{}
\caption{: Steps for generating the data set $u_s(\rvec,\tvec,\qvec(\tvec))$}
\label{array-sum}
\textbf{(1)} Fit a sphere or an ellipsoid to the realistic head model to estimate the centroid (centre). 
 The position vector of this centroid is be denoted by $\bold{c}$.\\
\textbf{(2)} Translate the coordinate system from $(0,0,0)$, i.e,  $\forall j$, let $\tvec_j=\bold{x}_j-\bold{c}$, $\rvec_j=\bold{y}_j-\bold{c}$.\\
\textbf{(3)} For all the nodes of the brain mesh, compute $\tvecunit_j:=\frac{\tvec_j}{|\tvec_j|}$. This is needed to define radially oriented sources.\\
\textbf{(4)} Find the minimum and maximum distance of the scalp nodes from the centroid $\bold{c}$, i.e., $r_{\textrm{min}}=\min\{|\rvec_j|:1\leq j\leq V_s\}$, $r_{\textrm{max}}=\max\{|\rvec_j|:1\leq j\leq V_s\}$.\\
\textbf{(5)}  Select $M$ nodes on the scalp (observation vectors) $\rvec_k$ in the interval $r \in [r_{\textrm{min}},r_{\textrm{max}}]$. \\
\textbf{(6)} It is well known that sources placed too close to the surface of cerebrum $S_c$, can result in inaccuracies in the numerical computation \cite{openmeeg2010,openmeeg2011}. Choose $V_b$  position vectors $3$ millimetres from each node of the brain mesh as source position vectors $\{\tvec^{s}_j:=\tvec_j-0.003\tvecunit_j:j\leq V_b\}$ and set the corresponding dipole moment for every node as $\qvec(\tvec_j)=\tvecunit_j$ to compute the corresponding electric potential, $\{u(\rvec_i,\tvec^{s}_j,\tvecunit_j): i\leq M, j\leq V_b\}$. \\
\textbf{(7)} Add $V_b$ sources at the centroid  $\bold{c}$, with the dipole moment $\qvec(\bold{c})=\tvecunit_j$, to compute $\{u(\rvec_i,\bold{c},\tvecunit_j): i\leq M, j\leq V_b\}$.    \\
\textbf{(8)} Add $V_b$ sources at the position vectors $\frac{\tau_j}{2}\tvecunit_j$ with dipole moments $\qvec(\frac{\tvec_j}{2})=\tvecunit_j$, to compute $\{u(\rvec_i,\frac{\tvec_j}{2},\tvecunit_j): i\leq M, j\leq V_b\}$.
\end{algorithm}

The above steps will result in a total of $M \times 3 V_b$ data points. For the purpose of constructing the surrogate model it is convenient to organize the data as a matrix $\bold{d}\in \Reals{3V_bM \times 4}$. The first three columns ($r,\tau,\tvecunit\cdot \rvecunit$) feature as inputs and the last column (electric potential) is the output. \par We employ the commercial software package pSeven by datadvance \cite{datadvance2018} to construct the regression model (machine learning model). A discussion of the features of pSeven and its comparison to open source software is presented in \cite{mikhail2016}. \par
A detailed and thorough investigation of machine-learning models (surrogate models) for the forward problem associated with EEG is outside the scope of this paper. It is work in progress and will be published elsewhere. It is noted that pSeven has a \textit{smart selection} option for scanning through a set of algorithms to select the best model amongst the set. It performs for each training set a numerical optimization of the technique as well as its parameters \cite{snoeck2012,bergstra2011} by minimizing the cross-validation error, see \cite{mikhail2016}. Among the algorithms scanned by pSeven are the following: ridge regression \cite{tikh1943}, stepwise regression \cite{efroymson1960}, elastic net \cite{zou2005}, Gaussian processes \cite{rasmussen2005}, sparse Gaussian processes \cite{candela2005,burnaev2015}, \textit{High Dimensional Approximation (HDA)} \cite{mikhail2016,haykin2008}, and \textit{High dimensional approximation combined with Gaussian processes (HDAGP)} (this technique is related to artificial neural networks and, more specifically, to the two-layer perceptron with a non-linear activation function\cite{haykin2008}). Two desirable features of pSeven are: (i) all data manipulation is done via graphical user interface (GUI) and (ii)  it can export the constructed surrogate model as a stand alone function in a number of scientific computing languages, including Matlab, C source for MEX, C source for stand alone program, C header for library, C source for library, functional mock-up interface (FMU) for Co-simulation $1.0$ and executable. 
\section{Minimization and a Numerical Solution of The Inverse Problem}
\label{inverseproblem}
It is clear from equations (\ref{2ndexpress2}) and (\ref{2ndexpress4}) that the only component of the neuronal current $\Jvec(\tvec)$ that affects EEG data is $\Psi(\tvec)$. However, as stated earlier, even this scalar function can not be computed uniquely, unless one imposes an appropriate constraint. It was shown in \cite{fokas1996,Kurylev2012} that for a spherical conductor the $L_2$  norm of $\Jvec(\tvec)$ (minimal energy) yields a unique solution. Here, these results are generalized to the case of a realistic head model. Let us define the functional $E$ (energy) by 
\begin{equation}
E=\int_{\Omega_c}|\Jvec|^2 dV(\tvec).
\label{energy}
\end{equation}
Using equation (\ref{helmddecomp}) we find that 
\begin{eqnarray}
&|\Jvec|^2=\biggl( \nabla \Psi+\nabla \times \Avec\biggr)\cdot \biggl( \nabla \Psi+\nabla \times \Avec\biggr).\\
&=|\nabla \Psi|^2+ |\nabla \times \Avec|^2+2 \nabla \Psi \cdot \nabla \times \Avec. \nonumber
\label{helmddecompsq}
\end{eqnarray}
Hence,
\begin{equation}
E=\int_{\Omega_c} \biggl(|\nabla \Psi|^2+ |\nabla \times \Avec|^2+2 \nabla \Psi \cdot \nabla \times \Avec\biggr) dV(\tvec). 
\label{energy2}
\end{equation}
However, it is shown in proposition $1$ and lemma $1$ of \cite{Cantarella2002} that 
\begin{equation}
\int_{\Omega_c} \nabla \Psi(\tvec) \cdot \nabla \times \Avec(\tvec) dV(\tvec)=0.
\end{equation} 
Thus, 
\begin{equation}
E=\int_{\Omega_c} \biggl(|\nabla \Psi|^2+ |\nabla \times \Avec|^2\biggr) dV(\tvec). 
\label{energy3}
\end{equation}
Taking into consideration that $u_s$ does not depend on $\Avec$ and that the second term of the right hand side of the above equation is positive, it follows that the minimization of (\ref{energy3}) is equivalent to minimizing 
\begin{equation*}
\int_{\Omega_c} |\nabla \Psi|^2 dV(\tvec). 
\end{equation*}
Thus, we have the following constrained minimization problem
\begin{eqnarray}      
\label{optproblem}
& \min_{\Psi(\tvec)} \int_{\Omega_c} |\nabla \Psi(\tvec)|^2 dV(\tvec)\\
& \mathrm{s.t.} \qquad u_s(\rvec)=-\frac{1}{4\pi} \int_{\Omega_c} \nabla_{\tvec}^{2} \Psi(\tvec) v_s(\rvec,\tvec)dV(\tvec), \qquad \rvec \in S_s \nonumber,
\end{eqnarray}
The requirement of the minimum norm of $\nabla \Psi(\tvec)$ is a well documented choice of regularization, known as Tikhonov regularization \cite{kaipio2004,tikh1943}. It forces the estimate $\Psi \in \Hilbert^1(\Omega_c)$. The continuum form of the constrained optimization problem given by (\ref{optproblem}) does not take into account the impact of  measurement noise on data $u_s(\rvec)$. However, additive measurement noise is modelled in section \ref{expPsiRBFsec}.

\subsection{Expansion of $\Psi(\tvec)$ Using Radial Basis Functions}
\label{expPsiRBFsec}
For the numerical minimization of the constrained optimization problem given in (\ref{optproblem}), radial basis functions (RBF) \cite{fornberg2015} are employed. We consider an EEG electrode cap with $M$ electrodes and discretize the domain $\Omega_c$ using $N$ cubic voxels. The function $\Psi(\tvec)$ is expanded using inverse multiquadric RBFs \cite{fornberg2015}, namely
\begin{equation}
\Psi(\tvec)=\sum_{j=1}^{N}\lambda_j \phi(\|\tvec-\tvec_j\|),
\label{psiparam}
\end{equation}
where $\{\tvec_j \in \Omega_c, 1\leq j\leq N\}$ are the position vectors of the centre of each voxel, and $\{\lambda_j:1\leq j\leq N\}$ denote the RBF coefficients. The inverse multiquadric basis function $\phi(\|\tvec-\tvec_j\|)$ is defined by
\begin{equation}
\phi(\|\tvec-\tvec_j\|):=\frac{1}{\sqrt{\alpha^2+\|\tvec-\tvec_j\|^2}}.
\label{mqbasis}
\end{equation}
The coefficient $\alpha$ in equation (\ref{mqbasis}) is referred to as the \textit{shape parameter} and needs to be estimated from appropriate data. In this study we propose a robust non-linear least squares strategy for estimating $\alpha$ from a data set  $\{u(\rvec_j):1\leq j\leq M\}$ generated by a known $\Psi(\tvec)$ which has support in $\Omega_c$.  In this setting, a convenient representation of equation (\ref{psiparam}) takes the linear algebraic form 
\begin{equation}
\Psiv= \mt{A}\lambdavec,
\label{psirbf}
\end{equation} 
where $\lambdavec \in \Reals{N}$ are the RBF coefficients. The entries of the matrix $\mt{A}\in \Reals{N \times N}$ are given by
\begin{equation}
\mt{A}[i,j]=\phi(\|\tvec_i-\tvec_j\|).
\label{Amatrix}
\end{equation}
If we assume that measurements are contaminated by additive noise, then the discretized form of equation (\ref{2ndexpress2}) is given by
\begin{equation}
\uvec=\mt{V_s} \PsiL+\wgn,
\label{uvecdiscrete1}
\end{equation}
where $\uvec[i]:=u(\rvec_i)$, $\mt{V_s}[i,j]:=-\frac{v_s(\rvec_i,\tvec_j)}{4\pi N}$, $\PsiL[j]:=\nabla^2 \Psi(\tvec_j),\quad j=1,...,N$ and $\wgn \in \Reals{M}$ denotes the measurement noise vector. The only unknown in equation (\ref{uvecdiscrete1}) are the discrete Laplacian values $\PsiL \in \Reals{N}$. The action of the Laplacian operator $\mt{L}$ on the inverse multiquadradic basis function of (\ref{mqbasis}) yields
\begin{equation}
\mt{L}[i,j]=-\frac{3\alpha^2}{(\alpha^2+\|\tvec_i-\tvec_j\|^2)^{\frac{5}{2}}}, \qquad 1\leq i,j\leq N.
\label{Lapop}
\end{equation}
The optimization problem defined in (\ref{optproblem}) requires the computation of $\nabla\Psi(\tvec)$.  Partial derivative operators enable $\nabla \Psi(\tvec)$ to be computed at discrete points $\{\tvec_j\}_{j=1}^N$. The operators for computing $\frac{\partial \Psi}{\partial \tau_x}$, $\frac{\partial \Psi}{\partial \tau_y}$, $\frac{\partial \Psi}{\partial \tau_z}$  are respectively given by 
\begin{equation}
\mt{D}_{x}[i,j]=- \frac{1}{\sqrt{N}}\frac{(\tau_{x,i}-\tau_{x,j})}{(\alpha^2+\|\tvec_i-\tvec_j\|^2)^{\frac{3}{2}}}, \quad i,j \leq N,
\label{Dx}
\end{equation}

\begin{equation}
\mt{D}_{y}[i,j]=-  \frac{1}{\sqrt{N}}\frac{(\tau_{y,i}-\tau_{y,j})}{(\alpha^2+\|\tvec_i-\tvec_j\|^2)^{\frac{3}{2}}}, \quad i,j \leq N,
\label{Dy}
\end{equation}

\begin{equation}
\mt{D}_{z}[i,j]=- \frac{1}{\sqrt{N}} \frac{(\tau_{z,i}-\tau_{z,j})}{(\alpha^2+\|\tvec_i-\tvec_j\|^2)^{\frac{3}{2}}}, \quad i,j \leq N.
\label{Dz}
\end{equation}

\subsection{Reconstructing $\Psi(\tvec)$}
\label{disreconPsisec}
The implementation of the constrained optimization problem given by (\ref{optproblem}) is well documented, see \cite{kaipio2004}. The shape parameter $\alpha$ that features in equations (\ref{Amatrix})-(\ref{Dz}) for computing the operators $\mt{A}$ $\mt{L}$, $\mt{D}_x$, $\mt{Dy}$ and $\mt{Dz}$ plays a central role in the accuracy and stability of reconstructions. In fact, the shape parameter $\alpha$ can be interpreted as the regularization parameter \cite{kaipio2004}. 
Extensive numerical tests suggest that the optimal $\alpha$ is not sensitive to the choice of $\Psi(\tvec)$ but it depends on the geometry of the head model, position vectors $\{\tvec_j \in \Reals{3}:j=1,...,N\}$.  Recall the Laplacian operator $\mt{L}$ given by equation (\ref{Lapop}), by letting $\PsiL=\mt{L} \lambda$ in equation (\ref{uvecdiscrete1}), we find 
\begin{equation}
\uvec=\mt{V_s} \mt{L}\lambdavec+\wgn,
\label{uvecdiscrete2}
\end{equation}
The model given by equation (\ref{uvecdiscrete2}) is linear in $\lambdavec \in \Reals{N}$, non-linear in the shape parameter $\alpha$, and importantly it is a separable model \cite{kay1992}. The least squares error may be minimized with respect to $\lambdavec$ and, thus it is reduced to a function of $\alpha$ only. The discretized form of the optimization problem of equation (\ref{optproblem}) is given by
\begin{eqnarray}      
\label{optproblem2}
& \qquad \min_{\alpha,\lambdavec} \|\mt{G}(\alpha) \lambdavec\|^2 \\
& \mathrm{s.t.} \quad \uvec=\mt{V_s} \mt{L}\lambdavec+\wgn \nonumber,
\end{eqnarray}
where $\|.\|^2$ is the  Euclidean norm and $\mt{G}(\alpha) \in \Reals{3N \times N}$ is defined as
\begin{equation*}
 \mt{G}(\alpha):=
  \left[ {\begin{array}{c}
   \mt{D_x}\\
   \mt{D_y}\\
   \mt{D_z}
  \end{array} } \right]
\end{equation*}
We employ the well documented weighted least squares method \cite{kay1992} to model the impact of the measurement noise $\wgn$ on the parameters $(\alpha,\lambdavec)$. The WLS method involves employing the covariance matrix of the measurement noise $\mt{C}_{\wgn}=\Exp[\wgn \wgn^{T}]$ in the cost function to be minimized. For convenience, we introduce the following notation :
 \begin{equation}
   \mt{H}(\alpha)=
  \left[ {\begin{array}{cc}
   \mt{C}^{\frac{1}{2}}_{\wgn}\mt{V_s}\mt{L} \\
   \mt{D_x}\\
   \mt{D_y}\\
   \mt{D_z}
  \end{array} } \right]
   \bvec=
  \left[ {\begin{array}{cc}
   \mt{C}_{\wgn}^{\frac{1}{2}} \uvec \\
   \zv\\
   \zv\\
   \zv
  \end{array} } \right],
\end{equation}
where $\mt{H}(\alpha) \in \Reals{(M+3N) \times N}$, $\zv \in \Reals{N}$ and $\bvec \in \Reals{M+3N}$. The optimization problem of equation (\ref{optproblem2}) is reduced to the following system of linear equations:
\begin{equation}
\label{optproblem3}
\bvec=\mt{H}(\alpha) \lambdavec.
\end{equation}
The dependency of $\mt{H}(\alpha)$ in equation (\ref{optproblem3}) on the parameter $\alpha$ is via the RBF matrices $\mt{L}$, $\mt{D}_x$, $\mt{D}_y$ and $\mt{D}_z$, respectively given by equations (\ref{Lapop})-(\ref{Dz}). The WLS error of equation (\ref{optproblem3}) reads as
\begin{equation}
J(\alpha,\lambdavec)=(\bvec-\mt{H}(\alpha) \lambdavec)^{T} (\bvec-\mt{H}(\alpha) \lambdavec).
\label{costlambda}
\end{equation}
The $\lambdavec \in \Reals{N}$ that minimizes equation (\ref{costlambda}) for a given $\alpha$ is given by
\begin{equation}
\lambdaest=(\mt{H}^T(\alpha) \mt{H}(\alpha))^{-1} \mt{H}^T(\alpha) \bvec.
\label{lambdaest}
\end{equation}
The resulting WLS error is given by
\begin{equation}
J(\alpha,\lambdaest)=\bvec^T\biggr[\bold{I}-\mt{H}(\alpha) \biggl( \mt{H}^{T}(\alpha) \mt{H}(\alpha)\biggr)^{-1} \mt{H}^{T}(\alpha) \biggr] \bvec,
\label{costFalpha2}
\end{equation}
where $\bold{I} \in \Reals{(M+3N) \times (M+3N)}$ is the identity matrix. Hence, the problem of finding the optimal value of the shape parameter $\alpha$ is reduced to the following 1-D maximization problem :
\begin{equation}
\alphaopt=\arg \max_{\alpha} \biggl[ \bvec^T \mt{H}(\alpha) \biggl( \mt{H}^{T}(\alpha) \mt{H}(\alpha)\biggr)^{-1} \mt{H}^{T}(\alpha) \bvec\biggr].
\label{alphaopteqn}
\end{equation}
The $\lambdavec$ that minimize equation (\ref{costlambda}) for the optimal shape parameter $\alphaopt$ is given by
\begin{equation}
\lambdaest=(\mt{H}^T(\alphaopt) \mt{H}(\alphaopt))^{-1} \mt{H}^T(\alphaopt) \bvec.
\label{lambdaoptimal}
\end{equation}
The optimal shape parameter $\alphaopt$ is employed in equation (\ref{Amatrix}) to compute the RBF matrix $\mt{A}$. Finally,  $\hPsi$ is given by
\begin{equation}
\hPsi=\mt{A}\lambdaest.
\label{Psiestimate}
\end{equation}

\section{Numerical Results}
\label{numerics}
In the interest of reproducibility and to encourage further research in this direction, the code (surrogate model and reconstruction routines), head model (triangulated meshes) and the electrode positions are made available via Github \url{https://github.com/parham1976}.  The realistic head model considered in this study, is taken from the sample data set in the OpenMEEG package. The number of nodes and triangles of the surface meshes  is shown in table \ref{nestedopenmeeg}. The centroid of the brain mesh is  $\bold{c}=[-0.0043,0.0169,0.0672]$ (this provides the center of the coordinate system for constructing the surrogate model). 
\begin{table}[H]
\begin{center}
\begin{tabular}{ |c||c|c|c|| }
 \hline
 Surface meshes & Number of nodes& Number of triangles \\
 \hline
 Cerebrum: $S_c$ &2562& 5120 \\
 \hline
 Skull $S_b$ &2562& 5120 \\
 \hline
 Scalp $S_c$ &2562& 5120 \\
 \hline
 \end{tabular}
 \end{center}
 \caption{This table shows the number of nodes and triangles in the realistic head model.}
 \label{nestedopenmeeg} 
 \end{table}

\subsection{Surrogate Model}
\begin{figure}[H]
 \includegraphics[scale=0.83]{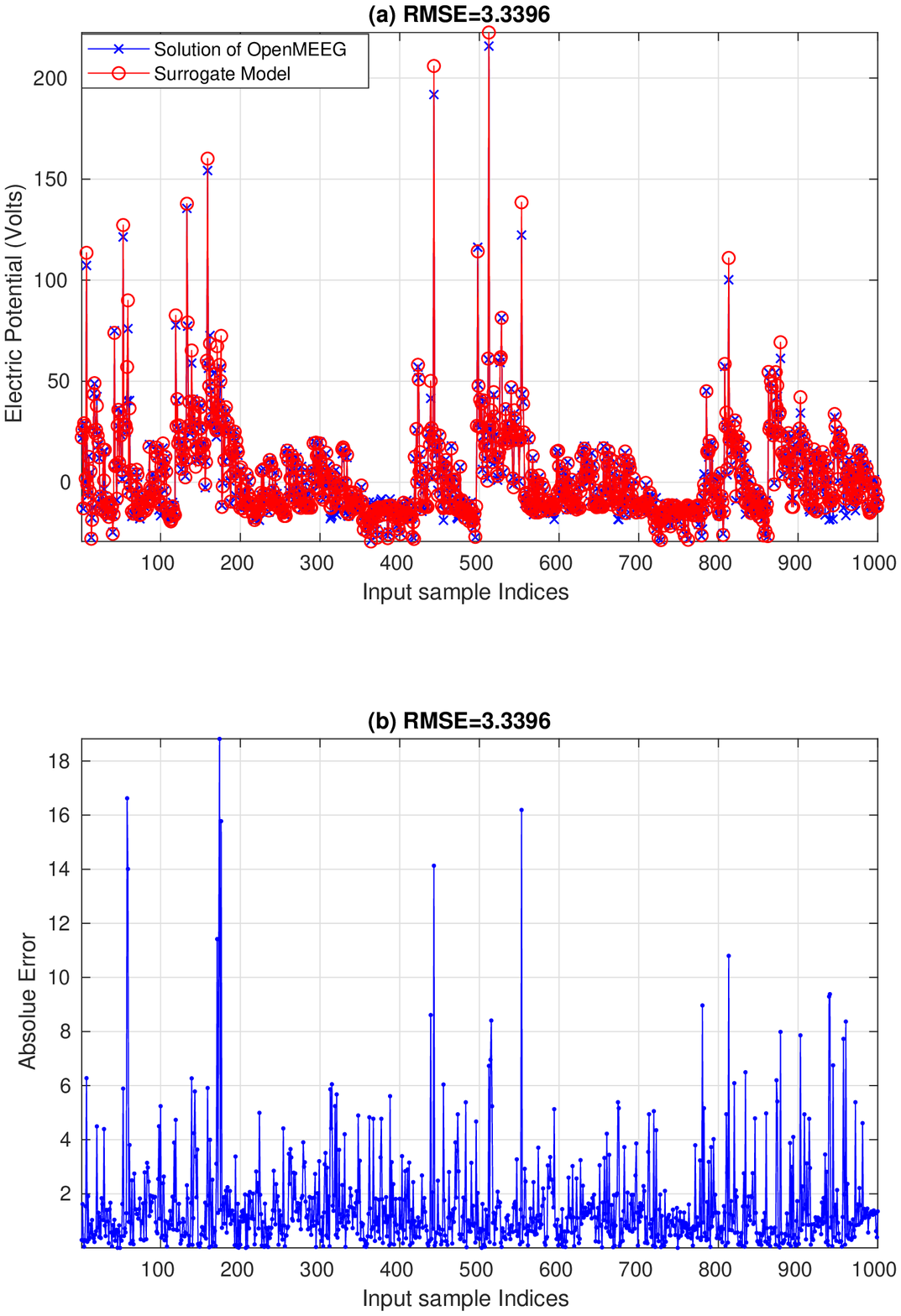}
  \caption{This figure depicts the comparison of the solution obtained via OpenMEEG(BEM)  and via the surrogate model on the test data set for $1000$ sample points. It shows the comparison of the solution of $\tvecunit_i \cdot \nabla_{\tvec} v_s(\rvec_i,\tvec_i)$ obtained via OpenMEEG and via the surrogate model. The solid blue line with cross is the OpenMEEG solution and the solid red line with circles is the surrogate model solution. The relative mean distance measure (RDM) is $0.2102$ and the natural logarithm of the magnification factor ($\ln$(MAG)) is $-0.0208$.}
  \label{crossvaldiationplot}
\end{figure}

In section \ref{forwardproblem}, the steps for generating the data set required for constructing the surrogate model were described. Those steps will yield $3V_bM$ data points. In this study, we have $122$ electrodes, i.e., $M=122$ and $V_b=2562$ as given in table \ref{nestedopenmeeg}. So, a total of $3V_bM=937692$ data points are generated via OpenMEEG. The data set is split into a training and a test data set. A total of $890807$ data points were randomly selected for training and the remaining $46885$ data points are used for testing. For the problem under study and the $890807$ samples chosens as training data, the \textit{smart selection} setting in pSeven package \cite{datadvance2018}, chose a two layer neural network. We employ the error metrics relative mean distance measure (RDM)$\in [0,2]$, where minimum RDM$=0$, as well as the natural logarithm of the magnification factor ($\ln$(MAG)) \cite{vorkwek2012,Reichenbach2010}. RDM is defined by 
\begin{equation}
\textrm{RDM}= \sqrt{\sum_{i=1}^n\biggl( \frac{u^{s}_i}{\sum_{j=1}^n (u^{s}_j)^2}-\frac{u^{o}_i}{\sum_{j=1}^n (u^{o}_j)^2}\biggr)^2},
\end{equation}
where $u^s_j$, $u^o_i$  respectively, denote the electric potential (Volts) estimated by the surrogate model and the OpenMEEG package (reference). The natural logarithm of the magnification factor is given by 
\begin{equation}
\ln[\textrm{MAG}]=\ln\biggl[ \sqrt{\frac{\sum_{i=1}^n (u^s_i)^2}{\sum_{i=1}(u^o_i)^2}}\biggr].
\end{equation}
Figure \ref{crossvaldiationplot} depicts the comparison of the solution obtained via OpenMEEG with the surrogate model. For clarity of presentation only $1000$ points from the test data set of $46885$ have been plotted. However, the values of RDM$=0.2102$  and $\ln$[MAG]$=-0.0208$, are computed using all $46885$ samples of the test data. In Figure \ref{crossvaldiationplot}, the blue solid line with cross is the solution obtained via OpenMEEG and the red solid line with circle is the output of the surrogate model. 

\subsection{Reconstruction of $\Psi(\tvec)$}
\label{numreconPsisec}
We present numerical results using synthetic data. In order to avoid the inverse crime \cite{kaipio2004}, we generate the data using a different model to that employed for reconstruction. We assume that $\Psi(\tvec)$ takes the form 
\begin{equation}
\Psi(\tvec)=\sum_{i=1}^4 \alpha_i e^{-\beta_i \|\tvec-\tvec_i\|^2},
\label{psiknown}
\end{equation}
where the values of $\{\alpha_i,\beta_i, \tvec_i:1\leq i\leq 4\}$ are given in the table \ref{tablpsiparam}.
\begin{table}[H]
\begin{center}
\begin{tabular}{ |c|c|c|c|c| }
 \hline
 $\alpha_i$ & $\beta_i$&$\tvec_{x,i}$&$\tvec_{y,i}$&$\tvec_{z,i}$\\
 \hline
  $2$&$1e3$&  $-0.0638$ &$-0.0185$ & $0.0546$\\
\hline
 $-2$&$1e3$&  -0.0424  &  0.0630  &  0.0324\\
\hline
 $2$&$1e3$&  0.0076 &  -0.0185  &  0.1139\\
\hline
 $-2$&$1e3$&  0.0433   &-0.0366 &   0.0398\\
\hline
 \end{tabular}
 \end{center}
 \caption{The table shows the values of the parameters $\{\alpha_i,\beta_i, \tvec_i:1\leq i\leq 4\}$ used for generating data.}
 \label{tablpsiparam} 
 \end{table}
The Laplacian of equation (\ref{psiknown}) is given by
\begin{equation}
\nabla^{2}\Psi(\tvec)=\sum_{i=1}^4 2\alpha_i \beta_i e^{-\beta_i\|\tvec-\tvec_i\|^2}  (-3+2 \beta_i \|\tvec-\tvec_i\|^2).
\label{LapPsiknown}
\end{equation}
The steps for generating the numerical results presented in this section are given below:
\begin{enumerate}
\item The head model will have a sensor array of $M=122$ sensors. Setup a uniform Cartesian grid in the domain $\Omega_c$. Discretize $\Omega_c$ using $N=3293$ voxels. The length of the edge for each voxel is $h=0.0074$ meters.
\item Compute and store $\mt{V_s}[i,j]:=-\frac{v_s(\rvec_i,\tvec_j)}{4\pi N}$. The auxiliary function matrix  $\mt{V_s} \in \Reals{M \times N}$.
\item Use in equation (\ref{LapPsiknown}) the values of $\{\alpha_i,\beta_i, \tvec_i:1\leq i\leq 4\}$ given in table \ref{tablpsiparam}  to compute 
$\{\nabla^2 \Psi(\tvec_j):j=1,...,N\}$. Use in equation (\ref{uvecdiscrete1}) $\mt{V_s}$ and $\PsiL[j]:=\{\nabla^2 \Psi(\tvec_j)\} $  to generate synthetic data, i.e., $\uvec=\mt{V_s}\PsiL+\wgn$. Here, we have employed additive white Gaussian noise (AWGN), with a signal to noise ratio of $20$ dB.
\item Solve the optimization problem of equation (\ref{alphaopteqn}) to find the optimal shape parameter, $\alphaopt$ and estimate $\hPsiL$. The optimal shape parameter is found to be $\alphaopt=0.0169$.
\item Use $\alphaopt$ to compute the linear operator $\mt{A}$ given by equation (\ref{Amatrix}). The reconstruction is given by equation (\ref{Psiestimate}), i.e., $\hPsi=\mt{A}\lambdaest$.
\end{enumerate}
The matrix of auxiliary function values $\mt{V}_s \in \Reals{122 \times 3293}$. It has a rank($\mt{V}_s$)=$122$ and condition number $\kappa(\mt{V_s})=7.9804 \times 10^{3}$. The matrix $\mt{H}(\alphaopt) \in \Reals{10001 \times 3293}$, which features in equation (\ref{optproblem2}) has a rank($\mt{H}(\alphaopt)$)$=3293$ and a condition number $\kappa(\mt{H}(\alphaopt))=8.4017 \times 10^{7}$. \par
\begin{figure}[H]
\includegraphics[scale=0.78]{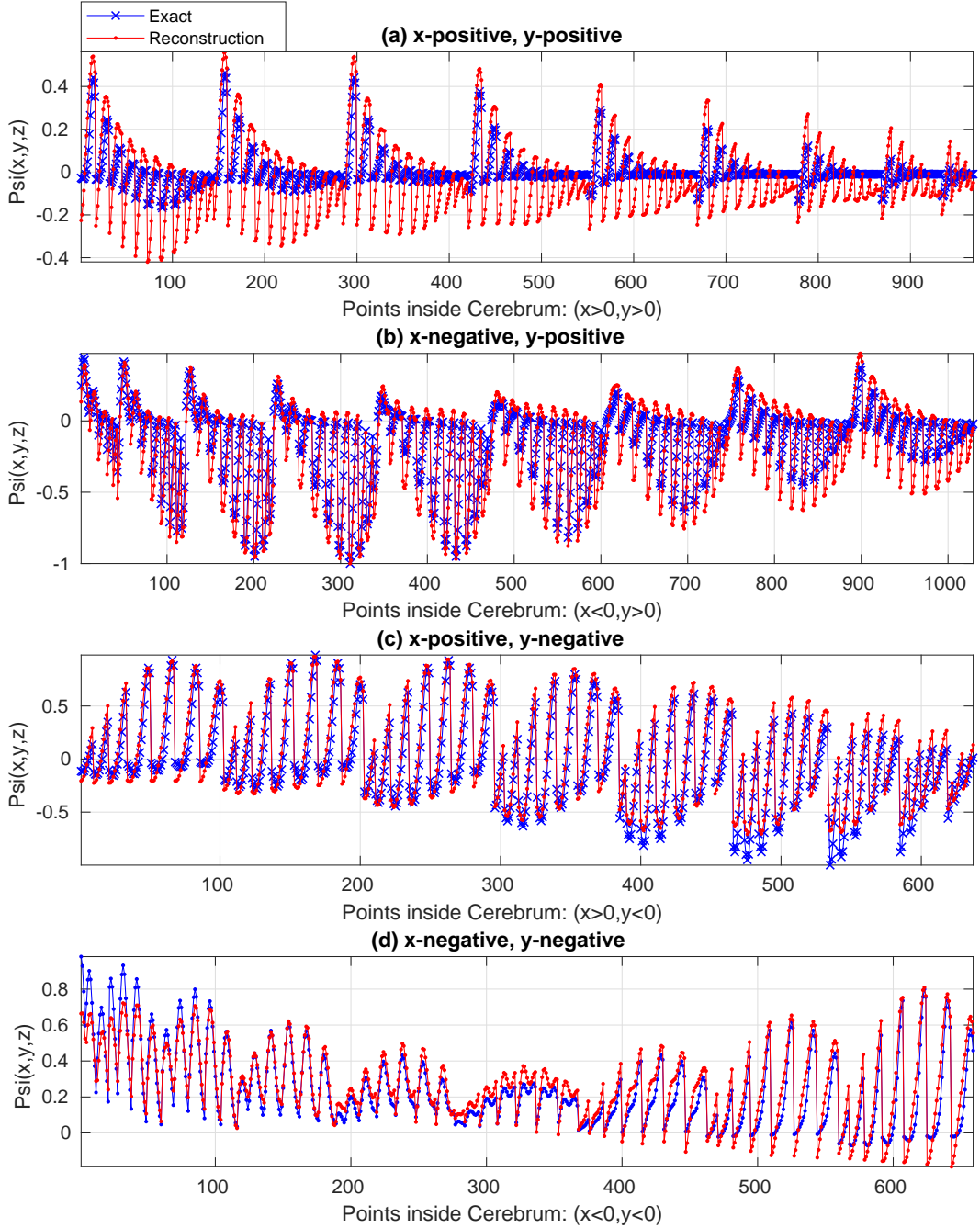}
\caption{This figure depicts the comparison of equations (\ref{psiknown}) and (\ref{Psiestimate}). The blue solid line with cross is $\{\Psi(\tvec_j): \tvec_j \in \Omega_c\}$ as given by equation (\ref{psiknown}) and the red solid line with dots is $\{\hPsi(\tvec_j):\tvec_j \in \Omega_c\}$ as given by equation (\ref{Psiestimate}). Subplot (a) depicts the comparison of the function values  of $\Psiv$ and $\hPsi$ for
the set of position vectors $\{\tvec_j \in \Omega_c: \tau_x>0, \tau_y>0\}$. Subplot (b) depicts the comparison of the function values of $\Psiv$ and $\hPsi$ for the set of position vectors $\{\tvec_j \in \Omega_c: \tau_x<0, \tau_y>0\}$. Subplot(c) depict the comparison of the function values of $\Psiv$ and $\hPsi$ for the set of position vectors $\{\tvec_j \in \Omega_c: \tau_x>0, \tau_y<0\}$. Subplot (d) depicts the comparison of the function values of $\Psiv$ and $\hPsi$ for the set of position vectors $\{\tvec_j \in \Omega_c: \tau_x<0, \tau_y<0\}$. The root mean square error (RMSE)=$0.1122$.}
\label{reconstruction1}
\end{figure}

Figures \ref{reconstruction1}(a)-\ref{reconstruction1}(d) depict the comparison between $\Psiv$ and $\hPsi$ for position vectors inside the domain $\{\tvec_j \in \Omega_c: 1\leq j\leq N\}$ (internal nodes). To facilitate visualization of the function values $\Psiv$ and $\hPsi$ inside the domain $\Omega_c$, the position vectors $\tvec_j \in \Omega_c$  are divided into four non-intersecting sets. The blue solid line with cross are the function values $\Psiv$ given by equation (\ref{psiknown}) and red solid line with dots is the estimate $\hPsi$ given by equation (\ref{Psiestimate}).  The domain $\Omega_c$ has been discretized using a uniform Cartesian grid, i.e., $N=3293$ cubic voxels. Subplot (a) depicts the comparison of the function values  of $\Psiv$ and $\hPsi$ for the set of position vectors $\{\tvec_j \in \Omega_c: \tau_x>0, \tau_y>0\}$. Subplot (b) depicts the comparison of the function values of $\Psiv$ and $\hPsi$ for the set of position vectors $\{\tvec_j \in \Omega_c: \tau_x<0, \tau_y>0\}$. Subplot(c) depicts the comparison of the function values of $\Psiv$ and $\hPsi$ for the set of position vectors $\{\tvec_j \in \Omega_c: \tau_x>0, \tau_y<0\}$. Finally, subplot (d) depicts the comparison of the function values of $\Psiv$ and $\hPsi$ for the set of position vectors $\{\tvec_j \in \Omega_c: \tau_x<0, \tau_y<0\}$. The root mean square error  (RMSE) of $\Psiv$ and $\hPsi$ is given by
\begin{equation}
\mathrm{RMSE}=\biggl[ \frac{1}{N} \sum_{i=1}^{N} \biggl( \Psi(\tvec_j)-\hat{\Psi}(\tvec_j)\biggr)^2 \biggr]^{\frac{1}{2}}.
\label{rmse}
\end{equation}
Using the analytic expression of $\Psi(\tvec)$ given by equation (\ref{psiknown}) and $\hPsi$ given by equation (\ref{Psiestimate}) in equation (\ref{rmse}), we find that RMSE=$0.1122$.

\begin{figure}[H]
\includegraphics[scale=0.6]{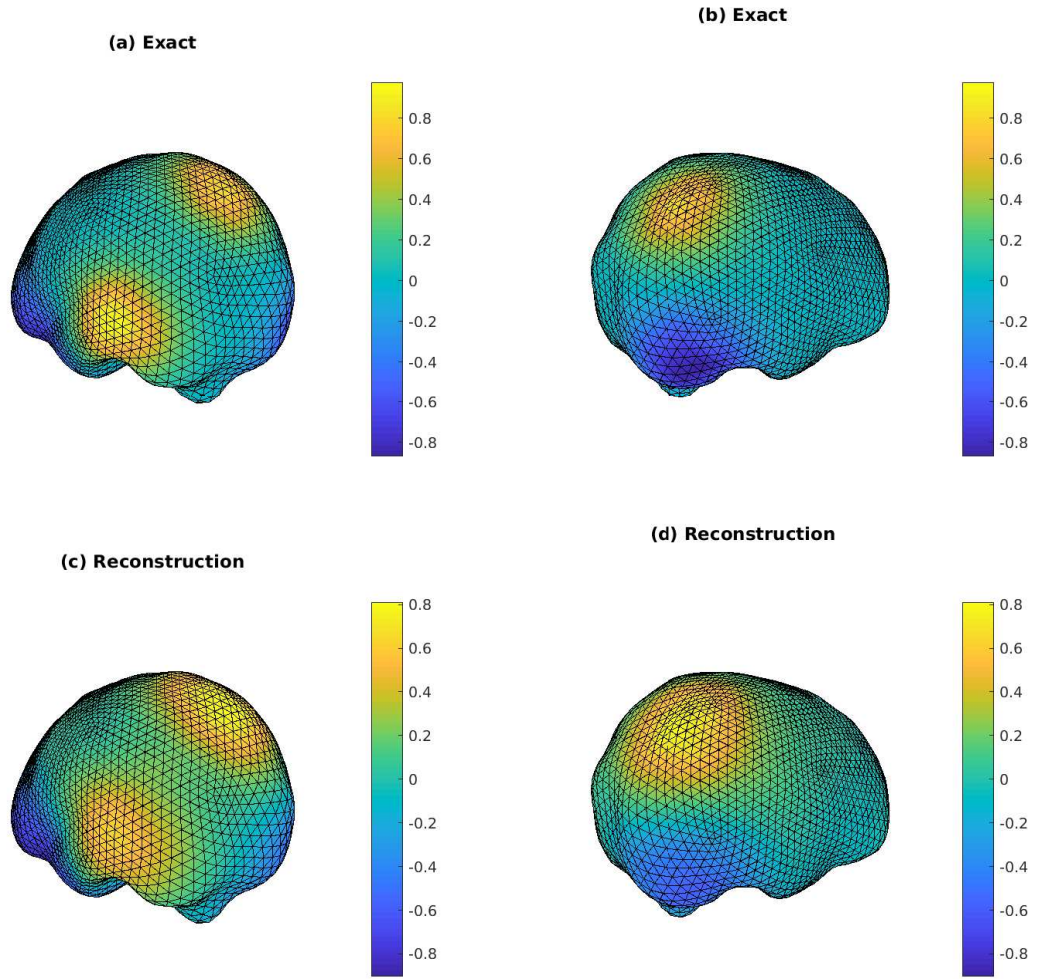}
\caption{This figure depicts the comparison of equations (\ref{psiknown}) and equation (\ref{Psiestimate}) on the surface of the cerebrum $S_c$. Subplots (a) and (b) depict $\Psiv$ on the surface $S_c$ as given by equation (\ref{psiknown}) and subplots (c) and (d) depict $\hPsi$ on $S_c$ as given by (\ref{Psiestimate}). The root mean square error (RMSE)=$0.1122$.}
\label{reconstruction2}
\end{figure}

Figure (\ref{reconstruction2}) depicts the comparison of $\Psiv$ as given by equation (\ref{psiknown}) on the surface $S_c$ with $\hPsi$ given by equation (\ref{Psiestimate}) on the surface $S_c$. 
The reconstructions in figure \ref{reconstruction2} are plotted using the open source fieldtrip package \cite{Fieldtrip}.
\section{Conclusion}
\label{conclusion}
OpenMEEG package solves the electric potential (Volts) $u_s(\rvec,\tvec)=\frac{1}{4\pi}\qvec(\tvec)\cdot \nabla_{\tvec} v_s(\rvec,\tvec)$ for a given dipole source  but not $v_s(\rvec,\tvec)$. To this end we propose a strategy involving 1-D line integrals to estimate $v_s(\rvec,\tvec)$ by constructing a surrogate model from data generated via OpenMEEG. The inputs to the surrogate model were selected based on insight provided by the analytic equation (\ref{vsansatz}), i.e., $v_s(\rvec,\tvec)$  for the spherical head model. The surrogate model was constructed using the pSeven Datadvance surrogate modelling toolbox \cite{datadvance2018}.
A total of $890807$ data points were randomly selected for training and the remaining $46885$ data points are used for testing. For this construction, the smart selection routine of Datadvance chose the high dimensional approximation \cite{mikhail2016}, which is essentially a two-layer neural network. The relative mean distance measure (RDM) is $0.2102$ and the natural logarithm of the magnification factor ($\ln$(MAG)) is $-0.0208$.  \par
The irrotational component of the neuronal current denoted by the scalar function $\Psi(\tvec)$ has been parametrized using inverse multiquadric radial basis functions (RBFs) on a uniform Cartesian grid inside the cerebrum of a realistic head model. By employing synthetic data, the shape parameter denoted by  $\alpha$ is estimated  using a computationally efficient approach and was estimated to be $\alphaopt=0.0169$. Furthermore, it is found that the shape parameter $\alpha$ is not sensitive to data, but depends on the configuration of the position vectors $\{\tvec_j \in \Omega_c: 1\leq j\leq N\}$ of the centres of the RBF expansion. The choice of RBF and $\alphaopt$ yields the condition number $\kappa(\mt{H}(\alphaopt))=8.4017 \times 10^{7}$ for the inversion matrix $\mt{H}(\alphaopt)$. Moreover, the regularization strategy involving minimizing the energy is in fact equivalent to Tikhonov regularization. Reconstructions are shown using synthetic data with a root mean square error (RMSE)=$0.1122$. The complete set of MATLAB files (surrogate model, inversion code and head model) as well as data set for reproducing the results in this paper are  available from \url{https://github.com/parham1976}.

\section*{Acknowledgement}
A.S. Fokas and P. Hashemzadeh are grateful to EPSRC and centre for mathematical imaging in healthcare (CMIH) for partial support. The grant number for this project is EP/N014588/1. C.B. Sch\"{o}nlieb acknowledges support from the EPSRC grant number EP/N014588/1. The authors are grateful for helpful comments, feedback and discussions with Martin Buhmann who is with the department of mathematics at the University of Giessen, as well as Jingwei Liang and No\'{e}mie Debroux who are with the Department of Applied Mathematics and Theoretical Physics, university of Cambridge.

\end{document}